# Time reversal symmetry breaking superconductivity in topological materials

Y. Qiu[1], K. N. Sanders[1], J. Dai[2], J. E. Medvedeva[1],

W. Wu[2], P. Ghaemi[3], T. Vojta[1] and Y. S. Hor[1]*

[1] Department of Physics, Missouri University of Science and Technology, Rolla, MO 65409

[2] Department of Physics and Astronomy, Rutgers University, Piscataway, NJ 08854

[3] Department of Physics, City College of the City University New York, New York, NY 10031

*Correspondence to: yhor@mst.edu

**Abstract**

Fascinating phenomena have been known to arise from the Dirac theory of relativistic quantum mechanics[1], which describes high energy particles having linear dispersion relations. Electrons in solids usually have non-relativistic dispersion relations but their quantum excitations can mimic relativistic effects. In topological insulators, electrons have both a linear dispersion relation, the Dirac behavior, on the surface and a non-relativistic energy dispersion in the bulk[2-6]. Topological phases of matter have attracted much interest, particularly broken-symmetry phases in topological insulator materials. Here, we report by Nb doping that the topological insulator $Bi_2Se_3$ can be turned into a bulk type-II superconductor while the Dirac surface dispersion in the normal state is preserved. A macroscopic magnetic ordering appears below the superconducting critical temperature of 3.2 K indicating a spontaneous spin rotation symmetry breaking of the Nb magnetic moments. Even though such a magnetic order may appear at the edge of the superconductor, it is mediated by superconductivity and presents a novel phase of matter which gives rise to a zero-field Hall effect.

Magnetism and superconductivity have been considered mutually exclusive since the 1950's when it was found that magnetic impurities strongly suppress superconductivity. Recent research has focused on unconventional superconductivity that can coexist with spin magnetism. Currently, only uranium compounds $UGe_2$[7], $URhGe$[8] and $UCoGe$[9] have been shown to exhibit superconductivity below 1 K inside the ferromagnetic phase, which is in proximity to a quantum phase transition. Coexistence of a magnetism and superconductivity in a topological insulator (TI) is still lacking though it has been theoretically postulated to be an important ingredient of odd-parity spin-triplet pairing for a topological superconductor (TSC). Similar to TIs, TSCs are predicted to have topologically protected edge excitations, provided by Majorana fermions instead of the proven Dirac fermions. Majorana fermions are considered exotic particles, which are identical to their own anti-particles[10]. Majorana fermionic bound state can serve as building blocks for fault tolerant topological quantum computing[11]; meanwhile, finding them in condensed matter is an active area of current research[12–14]. The TSC promising candidate is $Sr_2RuO_4$[15], which experiments suggest that it has a chiral p-wave spin-triplet superconductivity, analogous to the A phase of superfluid $^3He$. The Nb-intercalated $Bi_2Se_3$, $Nb_xBi_2Se_3$ becomes a superconductor. Although Nb is usually considered as non-magnetic cation, the intercalation also turns the parent compound $Bi_2Se_3$ into a magnetic system which indicates the possible presence of a p-wave spin-triplet superconductivity in $Nb_xBi_2Se_3$.

$Nb_xBi_2Se_3$ crystalizes into a rhombohedral structure, identical to that of $Bi_2Se_3$ as shown in Fig. 1(a) whereupon the $c$-parameter is slightly increased based on the powder x-ray diffraction of single crystals. First-principles density-functional calculations show that the intercalation of Nb in the van-der-Waals gap between the Se(1)-Bi-Se(2)-Bi-Se(1) quintuple layers is the most favorable. Moreover, only the intercalated Nb configuration gives a magnetic ground state. Its magnetic moment of 1.3 $\mu_B$ per Nb atom (obtained within the generalized gradient approximation, GGA + U, with a Coulomb repulsion of U = 1 eV for the Nb $d$-states) agrees well with the effective moment of 1.26 $\mu_B$/Nb obtained from the Curie law in Fig. 3(b). For comparison, our computational results show that intercalated Cu atoms in $Bi_2Se_3$ have a nonmagnetic ground state, which rests in agreement with the absence of magnetism in $Cu_xBi_2Se_3$[16–18].

Figures 1(b) and (c) show STM images taken on a cleaved Se surface of a $Nb_{0.25}Bi_2Se_3$ single crystal. The clean surface was obtained by in-situ cleaving under UHV condition (~ $10^{-11}$ torr) at room temperature. The images show tall protrusions with apparent heights of about 3 Å. The heights are bias-independent, suggesting that these features are formed from atoms above the cleaved surface. As these protrusions are absent in pure $Bi_2Se_3$, they likely correspond to intercalated Nb. Their lateral sizes vary, indicating that they are clusters of Nb atoms; however, the clustering may have happened after cleaving. The cloverleaf-shaped three-fold symmetric defects have been identified as $Bi_{Se}$ antisites[19,20]. The dark three-fold symmetric features shown in Fig. 1(b) can be attributed to the substitution of Bi with Nb but the substitutional Nb samples in the form of $Bi_{2-x}Nb_xSe_3$ do not superconduct at temperatures down to 2 K, which emphasizes the key role played by intercalating Nb atoms.

Dirac surface states are clearly visible in angle-resolved photoemission spectroscopy (ARPES) for both the $Bi_2Se_3$ TI and the $Nb_{0.25}Bi_2Se_3$ superconductor. Figure 2(b) shows that the Fermi level (FL) of the $Bi_2Se_3$ is in the bulk band gap, which differs from the ARPES data obtained by Xia *et al.*[21] because the $Bi_2Se_3$ crystal studied here was grown by our new method to reduce defects caused by the selenium vacancies[22]. In contrast, the FL of the $Nb_{0.25}Bi_2Se_3$ is located in the conduction band as both the bulk valence and conduction bands near the Γ point are clearly observed (Fig. 2(c)).

The longitudinal resistivity $\rho$ in Fig. 2(d) shows metallic behavior from room temperature to 3.6 K. The superconducting transition starts at 3.6 K and $\rho$ drops to zero at ~ 3.2 K (Fig. 2(e)). The field-dependent $\rho$ in Fig. 2(f) yields the upper critical fields $B_{c2\perp}$(2 K) = $\mu_0 H_{c2\perp}$ = 0.15 T for fields parallel to the crystallographic $c$-axis and $B_{c2\parallel}$ (2 K) = $\mu_0 H_{c2\parallel}$ = 0.31 T for fields parallel to the $ab$ plane. The temperature dependence of the upper critical fields is plotted in Fig. 2(g). Extrapolations to zero temperature give the values $B_{c2\perp}$(0 K) ≈ 0.2 T and $B_{c2\parallel}$(0 K) ≈ 0.4 T. This implies an in-plane coherence length of $\xi_{ab} = \sqrt{\Phi_0/(2\pi B_{c2\perp})}$ ≈ 40 nm. Analogously, from $\xi_{ab}\xi_c = \Phi_0/(2\pi B_{c2\parallel})$, we obtain $\xi_c$ ≈ 20 nm. We concluded that $Nb_{0.25}Bi_2Se_3$ is an anisotropic type-II superconductor with an anisotropic ratio of $B_{c2\parallel}/B_{c2\perp}$ ≈ 2. The total specific heat coefficient $c_p$ of a $Nb_{0.25}Bi_2Se_3$ multicrystalline sample showing an anomaly at the $T_c$ for a zero applied field demonstrates that the superconducting transition is a bulk transition (Fig. 2(h)). In 5 kOe field, the sample remains in the normal state down to the lowest temperatures. Its specific heat follows the conventional Debye fitting, $c_p = c_h + c_e = \gamma T + A_3 T^3 + A_5 T^5$, where $c_h$ and $c_e$ are phonon and electron contributions, respectively, and the normal-state electronic specific-heat coefficient $\gamma$ = 4.54 mJmol$^{-1}$K$^{-2}$.

Superconductivity of $Nb_xBi_2Se_3$ is verified by the Meissner effect shown in Fig. 3(a). It is observed at temperatures below 3.2 K with an applied AC field. The amplitude of the AC field is 10 Oe with zero offset. The AC magnetization is reversible, i.e., zero-field cooled (ZFC) and field-cooled (FC) magnetizations are overlapping. Superconducting volume fraction of the $Nb_{0.25}Bi_2Se_3$ is close to 100% which is much greater than that of $Cu_{0.12}Bi_2Se_3$[16] (~20%) indicating the full sample and not just local patches of Nb has become superconducting. In its normal state, $Nb_{0.25}Bi_2Se_3$ shows paramagnetic behavior as depicted in Fig. 3(b). The normal-state susceptibility and its inverse molar susceptibility of the sample follow Curie's law to give an effective moment of 1.26 µB/Nb. Figure 3(c) shows diamagnetic background-subtracted DC magnetization-field (MH) curves at $T$ = 2 K. For $H < H_{c2}$, the magnetization is anisotropic, and the MH curves feature the hysteresis loops of a strongly pinning type-II superconductor with unusual characteristics. As $H$ increases from zero as shown in the left inset of Fig. 3(c), the sample which initially has a zero-field magnetization (ZFM) starts to develop diamagnetism due to the Meissner effect. Increasing $H$ above a lower critical field $H_{c1}$ ($H_{c1\perp}$ ~ 23 Oe and $H_{c1\parallel}$ ~ 50 Oe) leads to a vortex state in which increasing magnetic flux lines penetrate the material. These vortex current-inducing fields in the vortices polarize Nb magnetic moments and cause a drastic increase from negative to positive magnetization in $H_{c1} < H < H_{c2}$. For $H > H_{c2}$, superconductivity is destroyed and the magnetic moments are ferromagnetically aligned by the external magnetic field. The magnetization of the normal state is isotropic and reversible. It shows saturation moment of ~ 0.3 $\mu_B$/Nb which is smaller than the fluctuating moment of 1.26 $\mu_B$/Nb determined from the inverse susceptibility in Fig. 3(b), suggesting an itinerant magnetism.

The MH curve for $H//c$ displays two unusual features. First, the $H//c$ initial MH curve shown in the left inset of Fig. 3(c) starts from a positive ZFM while the $H//ab$ initial MH curve shown in the right inset starts from a negative value. Second, the position of the maximum of the $H//c$ low-field MH curve or the central peak (CP) in the descending field branch is anomalously shifted to a positive field value, whereas the CP for the $H//ab$ low-field MH curve is at the negative field positon in the descending field branch. The negative CP position is normal and expected from the critical state models of strongly pinning type-II superconductors[23], but the positive CP position is unusual, which is ascribed to the internal magnetic field lagging behind the external one. The presence of the internal magnetic field is due to the mutual assistance between the induced vortex current and the magnetic moments. As shown in Fig. 3(c) for the $H//c$ magnetization, the magnetic moments become disoriented as $H$ decreases from the maximum causing a drop in the magnetization from the saturation moment to ~ 0.03 $\mu_B$/Nb at $H_{c2}$. For $H < H_{c2}$, superconducting vortices emerge to align back the magnetic moments restoring the positive magnetization to a maximum, about ¼ of the saturation moment near $H_{c1}$. Decreasing $H$ below $H_{c1}$ causes a sudden drop in the magnetization due to the vanishing of the mixed state Meissner effect but not at the edge. The edge magnetic state braced by the supercurrent persists even in the absence of external fields, which gives rise to the ZFM. The field at the edge produced by the magnetic state causes the shift of the CP, similar to what was seen in ferromagnet/superconductor/ferromagnet trilayered thin films[24].

Consistent with the ARPES, Hall effect measurements confirm the $n$-type conductivity at temperatures above $T_c$. In the normal state of $Nb_{0.25}Bi_2Se_3$, the transverse resistivity $\rho_{yx}$ at 3.8 K is a linear function of the applied magnetic field suggesting that a single type of electron carriers dominates the transport. The carrier density $n_e$, calculated from the Hall coefficient $R_H$ (Fig. 4(a)), is ~ $1.5 \times 10^{20}$ cm$^{-3}$. In the superconducting regime of $Nb_{0.25}Bi_2Se_3$, Hall resistivity is observed. The surprising properties of the superconducting phase of $Nb_xBi_2Se_3$, which are not observed in other doped TIs, indicate the effect of Nb dopants is beyond increasing the bulk career density. We believe that the sizable moment of Nb dopants in $Bi_2Se_3$ crystal, is crucially affecting the type of superconducting phase in this system. One of the striking features of $Nb_{0.25}Bi_2Se_3$ in this regards is the spontaneous Hall Effect which is present in superconducting phase (Fig. 4(b)) and provides a clear signature of time reversal symmetry breaking in the superconducting $Nb_{0.25}Bi_2Se_3$. Different mechanisms have been shown to lead to time reversal symmetry breaking in superconductors. One example in this regards was observation of Nernst signal in cuprates which has been attributed to the free energy gain in Josephson coupling between the layers by proliferation of same sign vortices. Such phenomenon is present for superconductors with pair-density waves[25]. The other example of time reversal symmetry broken superconductors, is the chiral p-wave triplet state which has been discussed extensively[26] in the context of $Sr_3RuO_4$ even though the experimental situation is not fully settled, yet. This state features topological order and Majorana edge states. Theoretical calculations show that impurity scattering in such chiral p-wave superconductor can indeed lead to a nonzero spontaneous Hall resistivity[27,28]. Given the strong spin orbit coupling in $Bi_2Se_3$, since the discovery of superconductivity in $Cu_xBi_2Se_3$ the possibility of a p-wave pairing in doped TIs was widely explored, but the commonly explored pairing was a time-reversal symmetric type of p-wave[29]. In contrary, our results strongly support a triplet chiral pairing in $Nb_{0.25}Bi_2Se_3$ which spontaneously breaks the time reversal symmetry. Even though settling the type of pairing in this system beyond doubt require further investigations, our results points to the importance of study of effect of dopants beyond mere change of chemical potential and opens a new venue of research in phases and phase transitions in doped topological insulators.

**Method**

A mixture of stoichiometric ratio of Nb (99.99%), Bi (99.999%) and Se (99.999+%) elements for $Nb_xBi_2Se_3$ was heated up to 1100 °C for 3 days in sealed vacuum quartz tubes. The crystal growth took place in cooling from 1100 °C to the quenched temperature 600 °C at 0.1 °C/min producing $Nb_xBi_2Se_3$ tiny crystals with sizes ranging from 0.1 to 1 mm. Omicron LT-STM under ultra-high vacuum (~ $1 \times 10^{-11}$ mbar) was used for the STM imaging and spectroscopic experiments at 4.5 K cooled by liquid helium. The differential conductance measurements were carried out using the standard lock-in technique with transimpedance amplifier gain $R_{gain} = 3$ G$\Omega$, AC modulation frequency $f = 455$ Hz and modulation amplitude $V_{mod} = 5$ mV. ARPES measurements were performed at the PGM beamline of the Synchrotron Radiation Center (SRC), Wisconsin using a Scienta R4000 analyzer with an energy resolution better than 5 meV and an angular resolution better than 0.2 ° (corresponding to a momentum resolution of $\Delta k \leq 0.009$ Å$^{-1}$ at photon energy $h\nu = 30$ eV). Physical properties were measured using a Quantum Design PPMS. Electrical resistivity measurements were made on a rectangular sample of uniform thickness (approximate size $1 \times 1 \times 0.1$ mm$^3$) using a standard four-probe technique. A small direct current of 3 mA was applied to prevent the sample from electrical heating. The Hall resistivity $\rho_{yx}$ was measured by a five-probe technique, where the electrical current was applied in the *ab* plane. The contacts for $\rho$ and $\rho_{yx}$ were made by attaching Pt wires with room-temperature cure silver paint. The specific-heat $c_p$ data were taken by a relaxation-time method using a commercial system; the addenda signal was measured before mounting the sample and was duly subtracted from the measured signal. The $c_p$ measurement was done in zero field for the superconducting state and 5 kOe was applied along the *c* axis for the normal state, while the change of the addenda signal between the two was found to be negligible. The AC and DC magnetizations were measured with the PPMS.

**Acknowledgements**

We gratefully acknowledge D. Singh, T. Hughes, C. Kurter, Lu Li and Liang Fu for important suggestions. Thanks to SRC for providing the beamline for ARPES experiment. This work was supported by the National Science Foundation (NSF) grant DMR-1255607. The work at Rutgers was supported by the NSF grant DMR-0844807. The work was also supported in part by the NSF under grant DMR-1205803. PG acknowledges partial support from NSF **EFRI**-1542863.

**Author contributions**

YSH conceived and directed the project from conception to production. YQ and YSH performed the crystal growth and structural characterizations, transport and magnetization measurements, and data analysis. YSH conducted the ARPES experiment. KNS and JEM carried out the computational studies of the structure and magnetic properties. STM measurement was done by JD and WW. TV and PG provided theoretical contribution. All authors contributed to the writing of the final version of the manuscript.

**Figure 1.** $Nb_xBi_2Se_3$ crystal structure. (a) Crystal structure of $Nb_xBi_2Se_3$ with a favorable intercalating Nb (the purple circle) site. Light blue and dark green solid circles are Bi and Se respectively. STM topographic images on a vacuum-cleaved surface of $Nb_{0.25}Bi_2Se_3$ at bias voltages of −0.4 and +0.5 V are shown in (b) and (c), respectively. The black circles mark the protrusions appearing in both images identified as Nb on the cleaved surface.

**Figure 2.** Dirac Surface State of $Nb_{0.25}Bi_2Se_3$. (a) A schematic diagram of bulk 3D Brillouin zone (BZ) of $Bi_2Se_3$ crystals and the 2D BZ of the projected (111) surface. (b) ARPES data taken at 12.8 K for a $Bi_2Se_3$ single crystal grown by the two-step method. It shows the surface state with the Dirac point ~0.15 eV below the FL. The FL lies in the band gap indicating that the $Bi_2Se_3$ is a nearly-perfect topological insulator. (c) FL in $Nb_{0.25}Bi_2Se_3$ lies in the bulk conduction band indicating an *n*-type conductor. The Dirac point is ~0.3 eV below the FL. (d) Temperature-dependent resistivity of $Nb_{0.25}Bi_2Se_3$ at $H = 0$ shows metallic behavior in its normal state identical to that of $Bi_2Se_3$. (e) The resistivity starting to drop at 3.6 K and becoming zero at 3.2 K. (f) Magnetic-field dependent resistivity of the $Nb_{0.25}Bi_2Se_3$ at 2 K for *H//c* and for *H//ab*. (g) Upper critical field $H_{c2}$ as a function of temperature for *H//c* and *H//ab*. $H_{c2}$ is defined by 50% of the normal-state resistivity $\rho(H_{c2}) = 0.5\rho(T_c)$. (h) Temperature dependence of specific heat measured in 0 and 5 kOe applied fields. The dashed line is the Debye fitting to the 5 kOe data.

**Figure 3.** Magnetic properties of $Nb_{0.25}Bi_2Se_3$. (a) AC magnetization as a function of temperature for a sample consisting of 15 tiny $Nb_{0.25}Bi_2Se_3$ single crystals. The crystals are stacking up with their *c*-axes oriented parallel to each other. *H* is parallel to the *c*-axes of the crystals (*H//c*). (b) Temperature-dependent magnetic susceptibility (red circles) of the same sample at 10 kOe DC applied magnetic field *H*. The susceptibility is diamagnetic background-subtracted which is $\chi - \chi_0$. The inverse molar susceptibility (green circles) of the sample shows

paramagnetic behavior that follows Curie's law. (c) Field-dependent DC magnetization (MH) for $H//c$ (red circles) and $H//ab$ (blue triangles) at 2 K. The left inset (right inset) shows the low-field region of the $H//c$ ($H//ab$) MH curve. Arrows show ascending and descending field branches.

**Figure 4:** Zero-field Hall effect in superconducting regime. (a) Hall resistivity for $Nb_{0.25}Bi_2Se_3$ single crystal. The applied magnetic field is along $c$-axis of the single crystal and the current is along the crystal $ab$-plane. (b) The zero-field Hall resistivity $\rho^0_{yx}$, the amplitude of the Hall resistivity in (a) at $H = 0$, is plotted as a function of temperature.

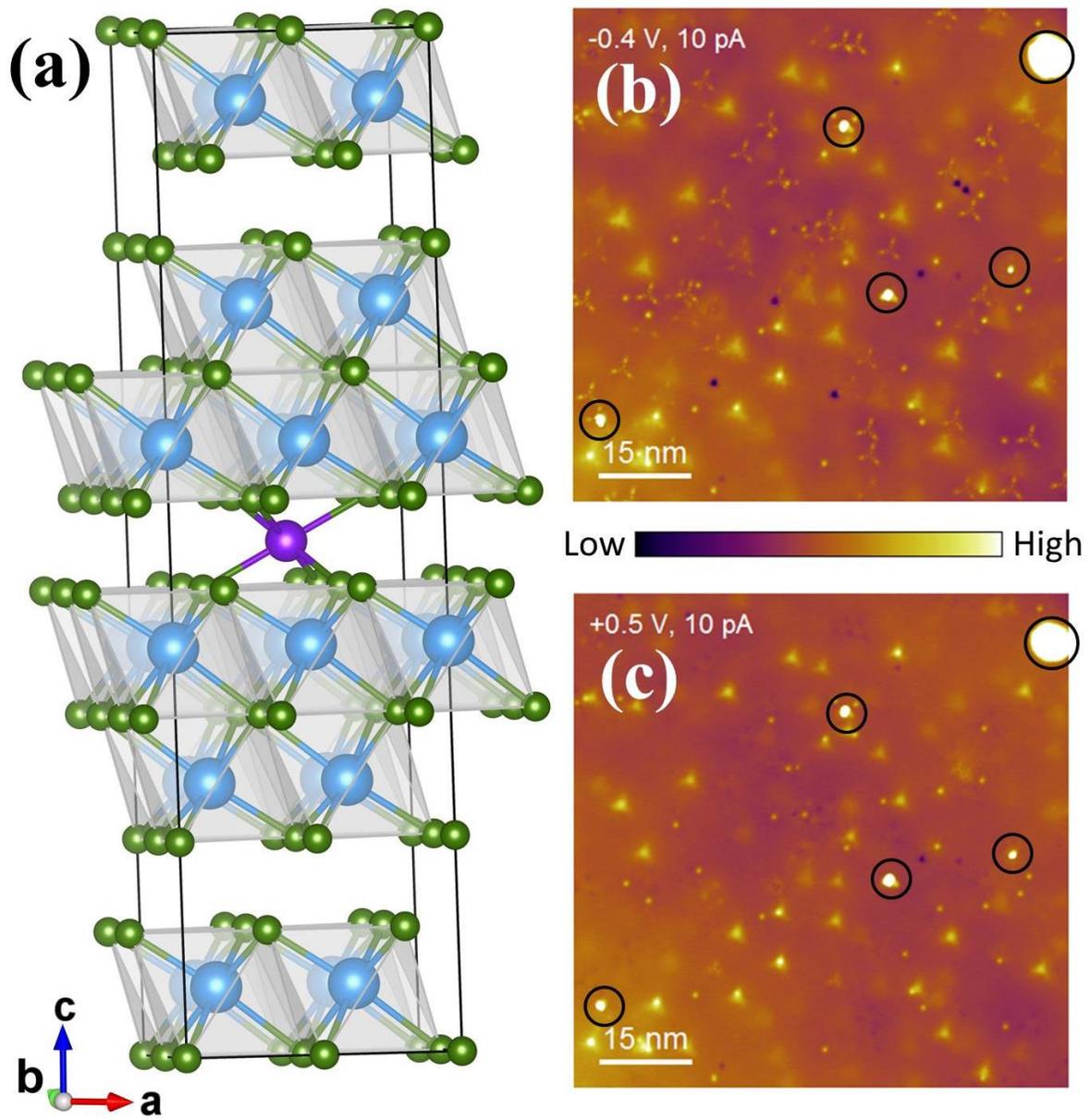

**Figure 1**

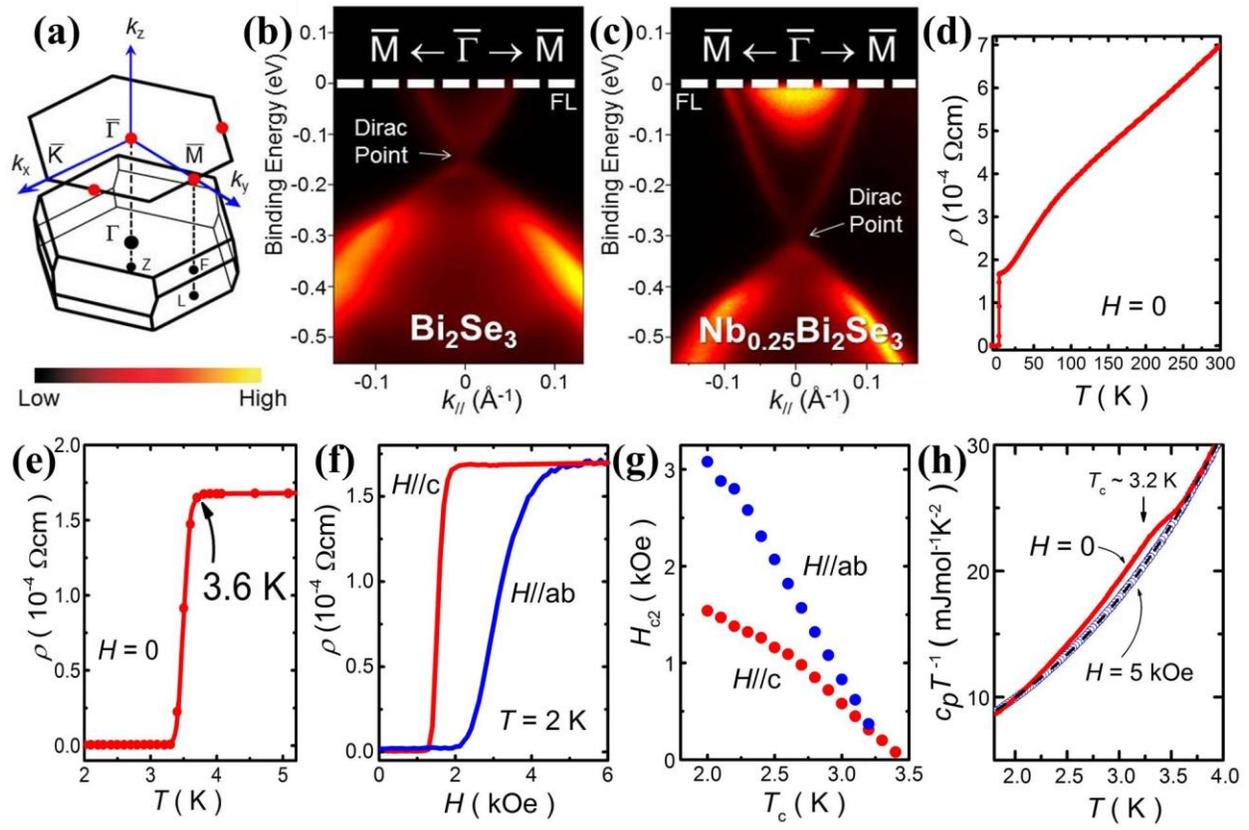

**Figure 2**

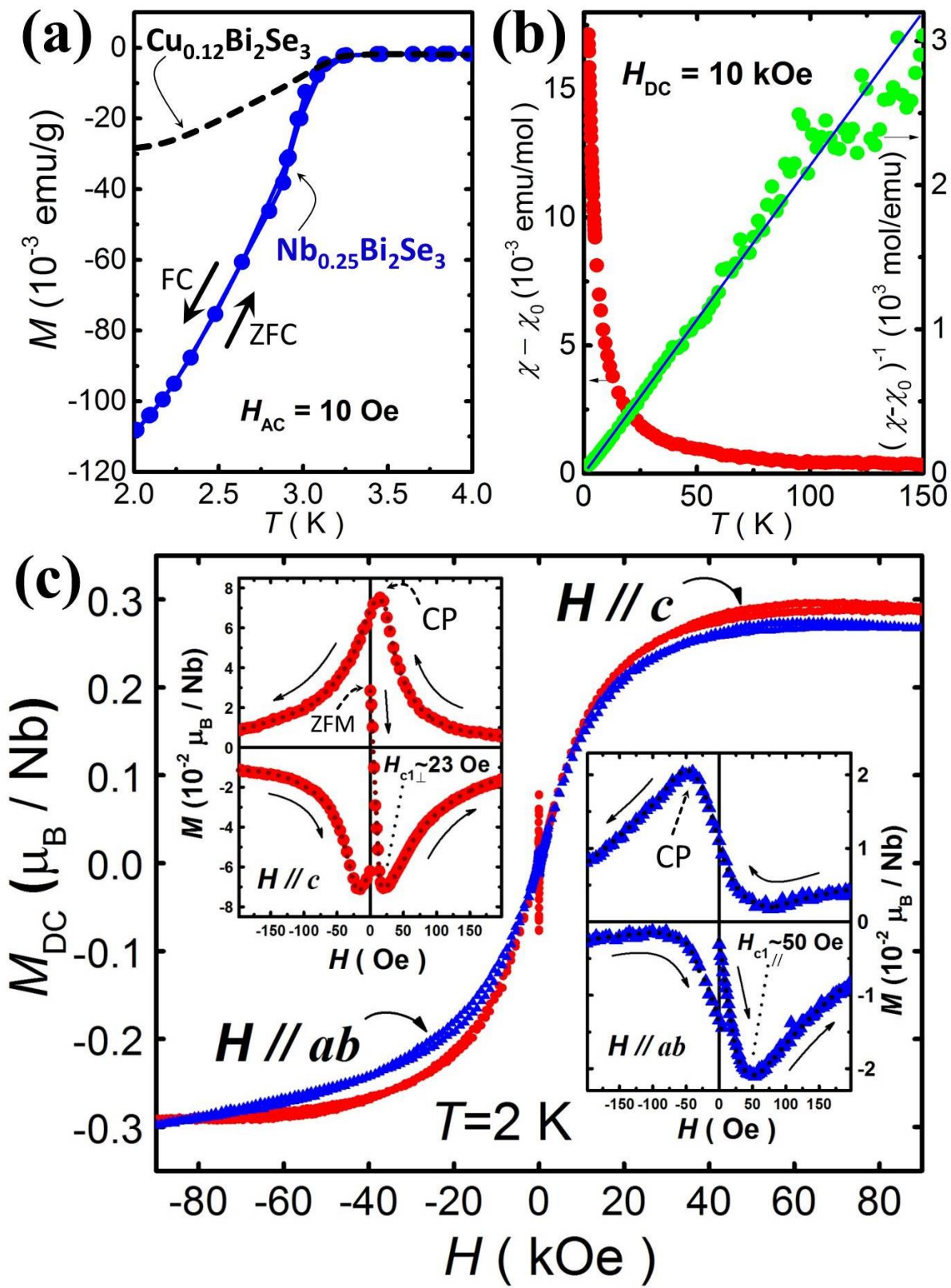

**Figure 3**

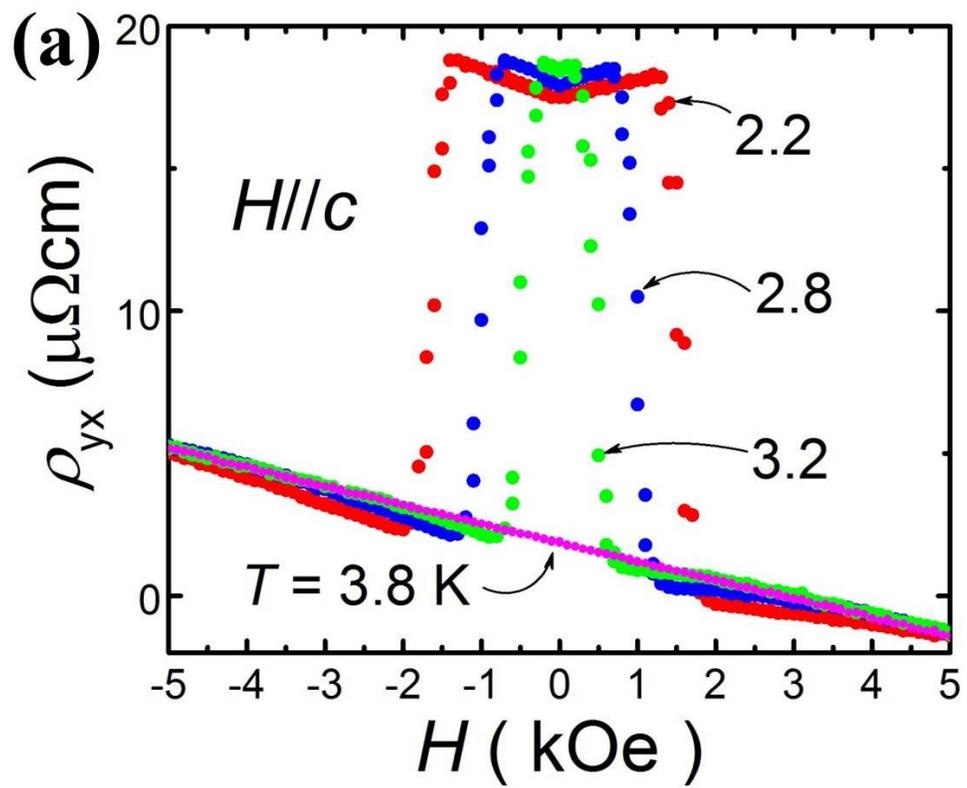

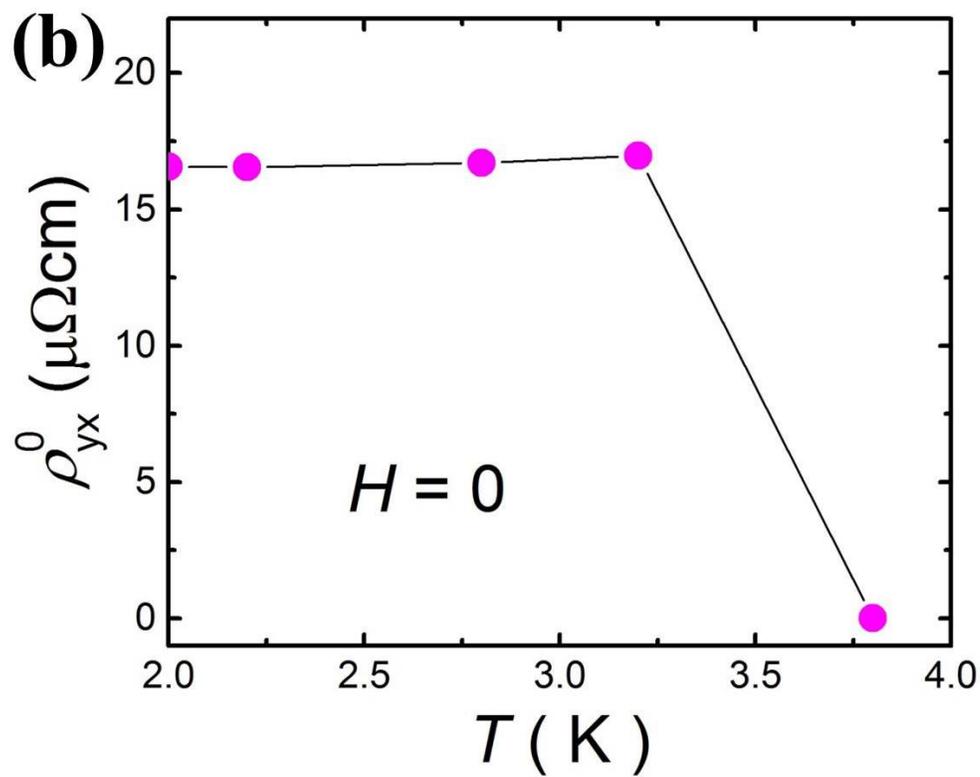

**Figure 4**